\documentclass[a4paper,prd,preprint,amsmath,
               superscriptaddress,nofootinbib,floatfix,showpacs]{revtex4}
\usepackage{graphicx}

\DeclareMathOperator{\im}{Im}

\DeclareMathOperator{\br}{B}

\newcommand{\bb}{\ensuremath{B^0-\bar{B}^0}}
\newcommand{\kk}{\ensuremath{K^0-\bar{K}^0}}
\newcommand{\bdbd}{\ensuremath{B_d-\bar{B}_d}}
\newcommand{\bsbs}{\ensuremath{B_s-\bar{B}_s}}
\newcommand{\acp}{\ensuremath{A_{\rm CP}^{\rm mix}(B\to J/\psi\,K_S)}}

\newcommand{\bsg}{\ensuremath{b\to s\,\gamma}}
\newcommand{\meg}{\ensuremath{\mu\to e\,\gamma}}
\newcommand{\ek}{\ensuremath{\varepsilon_K}}
\newcommand{\dmbd}{\ensuremath{\Delta m_{B_d}}}
\newcommand{\dmbs}{\ensuremath{\Delta m_{B_s}}}
\newcommand{\dmbsd}{\ensuremath{\Delta m_{B_s}/\Delta m_{B_d}}}
\newcommand{\e}{{\rm e}}

\begin{document}

\title{Exploring Flavor Structure of Supersymmetry Breaking\\ 
       at B factories}

\date{6/3/2002}

\author{Toru Goto}
\altaffiliation[Address after April 1, 2002: ]{Department of Physics,
Graduate School of Science,  Osaka University, Toyonaka, Osaka 560-0043, Japan}
\email{goto@het.phys.sci.osaka-u.ac.jp}
\affiliation{YITP, Kyoto University, Kyoto 606-8502, Japan}

\author{Yasuhiro Okada}
\email{yasuhiro.okada@kek.jp}
\affiliation{Theory Group, KEK, Tsukuba, Ibaraki 305-0801, Japan}
\affiliation{Department of Particle and Nuclear Physics,
             The Graduate University of Advanced Studies,
             Tsukuba, Ibaraki 305-0801, Japan}

\author{Yasuhiro~Shimizu}
\email{shimizu@eken.phys.nagoya-u.ac.jp}
\affiliation{Department of Physics, Nagoya University, Nagoya 464-8602, Japan}

\author{Tetsuo Shindou}
\email{shindou@het.phys.sci.osaka-u.ac.jp}
\author{Minoru Tanaka}
\email{tanaka@phys.sci.osaka-u.ac.jp}
\affiliation{Department of Physics, Graduate School of Science, 
             Osaka University, Toyonaka, Osaka 560-0043, Japan}

\begin{abstract}
We investigate quark flavor signals in three different supersymmetric
models, the minimal supergravity, the SU(5) SUSY GUT with right handed
neutrinos, and the minimal supersymmetric standard model with U(2) flavor 
symmetry, in order to study physics potential of the present and future 
$B$ factories. We evaluate CP asymmetries in various $B$ decay modes,
$\Delta m_{B_s}$, $\Delta m_{B_d}$, and $\varepsilon_K$. The allowed
regions of the CP asymmetry in $B\rightarrow J/\psi\, K_S$ and
$\Delta m_{B_s}/\Delta m_{B_d}$ are different
for the three models so that precise determinations of these observables
in near future experiments are useful to distinguish the three models.
We also investigate possible deviations from the standard model
predictions of CP asymmetries in other $B$ decay modes.
In particular, a large deviation is possible for the U(2) model.
The consistency check of the unitarity triangle including
$B\rightarrow \pi\pi,\rho\pi,D^{(*)}K^{(*)},D^{(*)}\pi,D^{*}\rho$,
and so on, at future high luminosity $e^+e^-$ $B$ factories and 
hadronic $B$ experiments is therefore important to distinguish 
flavor structures of different supersymmetric models.
\end{abstract}


\preprint{YITP-02-22}
\preprint{KEK-TH-815}
\preprint{DPNU-02-06}
\preprint{OU-HET 408}
\preprint{hep-ph/0204081}

\maketitle

%%%%%%%%%%%%%%%%%%%%%%%%%%%%%%%%%%%%%%%%%%%%%%%%%%%%%%
%
%
%	Introduction 
%	Distinguishing SUSY models at high luminosity B factory
%
%%%%%%%%%%%%%%%%%%%%%%%%%%%%%%%%%%%%%%%%%%%%%%%%%%%%%%
%%%%%%%%%%%%%%%%%%%%%%%%%%%%%%%%%%%%%%%%%%%%%%%%%%%%%
%	Introduction
%%%%%%%%%%%%%%%%%%%%%%%%%%%%%%%%%%%%%%%%%%%%%%%%%%%%%
\section{Introduction}
Recent development in $B$ physics is remarkable. 
Both the Belle experiment at KEK and the BaBar experiment 
at SLAC observed large CP violation in $B \rightarrow
J/\psi\, K_S$ and related modes. These observations are
the first discovery of CP violation out of 
the kaon system\cite{Abashian:2001pa,Aubert:2001sp}.
The results are consistent with the Kobayashi-Maskawa 
mechanism\cite{Kobayashi:1973fv} of CP violation 
in the three-generation Standard Model (SM).
In the coming years, we expect much improvements in 
measurements of CP violation and rare decay processes at the 
asymmetric $B$ factories. In addition, the magnitude of
the $B_s-\bar B_s$ mixing will be determined at the Tevatron
experiments\cite{Anikeev:2001rk}. 
It will be clear in a few years whether or not
the Cabibbo-Kobayashi-Maskawa (CKM) matrix is the main source of
the flavor mixing and the CP violation in the quark sector.

In future, $B$ physics is expected to play an even more important 
role in precisely determining the flavor structure of the SM
and searching for possible new physics effects beyond the SM. 
LHC-B\cite{LHC-B LOI} and BTeV\cite{Kulyavtsev:2000}
experiments are planned to provide very precise
information on the angles of the unitarity triangle from $B_d$ and
$B_s$ decays at hadron machines. As for $e^+e^-$ asymmetric colliders,
both KEK and SLAC are considering to increase luminosity by
$1-2$ orders of magnitude by the time that these hadron experiments
will be carried out\cite{Abe:2002aa,Zhao:2002zk}.
With luminosity of $10^{35} -10^{36} \text{cm}^{-2} \text{s}^{-1}$,
the super $B$ factories will provide us with $10^9 - 10^{10}$ $B\bar{B}$ 
pairs in a year.  Then, we shall have good opportunities to explore 
new physics from observations of CP asymmetries 
and rare $B$ decay processes.

Among various candidates of new physics beyond the SM,
supersymmetry (SUSY) is the most interesting one. Although the main 
motivation for introducing SUSY is to solve the hierarchy problem, namely
to give a justification to the electroweak scale, which is much smaller
than the Planck scale, flavor physics can provide important information
on SUSY models.
In SUSY models, mass matrices of SUSY partners 
of the quarks and the leptons are new sources of the flavor mixing.
Since these mass matrices are determined from SUSY breaking terms 
in the Lagrangian, 
their flavor structures reflect the SUSY breaking mechanism
and interactions present between the scale of the SUSY breaking 
and the electroweak scale.  
Future $B$ physics is therefore very important to discriminate
various SUSY breaking scenarios. It can play a more important role
if LHC experiments discover SUSY particles, in which case 
we can make more precise predictions for flavor signals based on
a particular scenario.

In this paper, we investigate SUSY effects on $B$ physics 
based on three different SUSY models, namely
(1) the minimal supergravity model (mSUGRA),
(2) the SU(5) SUSY GUT with right-handed neutrinos,
and (3) the SUSY model with U(2) flavor
symmetry\cite{Pomarol:1996xc,Barbieri:1997ww}.
We focus on the new physics search through
the consistency test of the unitarity triangle. We address two
questions. First, we ask  whether these models can be distinguished 
from the SM in near future by measuring the CP asymmetry of     
$B \rightarrow J/\psi\, K_S$ and the $B_s-\bar B_s$ mixing. Second, 
we consider impacts of other angle measurements of the unitarity
triangle  in the era of LHC-B/BTeV and an $e^+e^-$ super $B$ factory
when the above two observables are precisely determined.
We analyze the three models in the same fashion, so that we can clearly 
show  the potential of future $B$ physics. There are already many
analyses in the 
literature for each of three models\cite{Pomarol:1996xc,Barbieri:1997ww,
Bertolini:1991if,Baek:2001sj,Moroi:2000,Baek:2001kh,Masiero:2001cc},
but here we make a systematic
treatment to show that various
measurements in $B$ physics are 
in fact useful to distinguish  different SUSY models.
We show that the allowed region of $\Delta m_{B_s}/\Delta m_{B_d}$ is quite 
different from the SM prediction for the SU(5) SUSY GUT with
right-handed neutrinos and the U(2) model, whereas the deviation
is small for the mSUGRA. Furthermore, the GUT and the U(2) model 
can be distinguished when we measure  CP asymmetries of decay modes
such as $B\rightarrow \pi\pi,\rho\pi,D^{(*)}K^{(*)},D^{(*)}\pi$, 
and $D^{*}\rho$.

This paper is organized as follows. The three models are introduced in
Sec.~\ref{MODELS}. The $B_d-\bar B_d$ mixing, the $B_s-\bar B_s$
mixing, the CP violating parameter in the $K^0-\bar{K}^0$ mixing (\ek), 
and CP violations in various $B$ decays are discussed
in Sec.~\ref{OBSERVABLES}. The numerical results of these observables
are presented in Sec.~\ref{NR}. Our conclusion is given in
Sec.~\ref{CONCLUSIONS}.

%%%%%%%%%%%%%%%%%%%%%%%%%%%%%%%%%%%%%%%%%%%%%%
% Section 2 of the paper by T.Goto, Y.Okada, %
% Y.Shimizu, T.Shindou and M.Tanaka.         %
% Written by M.T. 2002/1/15-                 %
%%%%%%%%%%%%%%%%%%%%%%%%%%%%%%%%%%%%%%%%%%%%%%
\section{Models\label{MODELS}}
\subsection{The Minimal Supersymmetric Standard Model}
The minimal supersymmetric standard model (MSSM) is an 
$\text{SU(3)}_{\text{C}}\times \text{SU(2)}_{\text{L}}\times 
 \text{U(1)}_{\text{Y}}$ supersymmetric gauge theory
with the SUSY being softly broken. The MSSM matter
contents are the following chiral superfields:
\begin{eqnarray}
  &&Q_i  (3           ,\,2,\, \frac{1}{6}) ~,~~
    \overline{U}_i  (\overline{3},\,1,\,-\frac{2}{3}) ~,~~
    \overline{D}_i  (\overline{3},\,1,\, \frac{1}{3}) ~,
\nonumber\\
  &&L_i  (1           ,\,2,\,-\frac{1}{2}) ~,~~
    \overline{E}_i  (1           ,\,1,\, 1          ) ~,
\nonumber\\
  &&H_1  (1           ,\,2,\,-\frac{1}{2}) ~,~~
    H_2  (1           ,\,2,\, \frac{1}{2}) ~,
  \label{MSSMsuperfields}
\end{eqnarray}
where the gauge quantum numbers are shown in parentheses and
$i=1,2,3$ is a generation index. Assuming R-parity invariance
and renormalizability, we can write the MSSM superpotential  as%
\begin{equation}
  \mathcal{W}_{\text{MSSM}}
  = f_D^{ij} \overline{D}_i Q_j H_1
    + f_U^{ij} \overline{U}_i Q_j H_2
    + f_E^{ij} \overline{E}_i L_j H_1
    + \mu H_1 H_2 ~.
  \label{MSSMsuperpotential}
\end{equation}
The soft SUSY breaking is described by
the following Lagrangian:
\begin{eqnarray}
  -\mathcal{L}_{\text{soft}}
  &=& (m_Q^2)^i_{~j} \widetilde{q}_i \widetilde{q}^{\dagger j}
    + (m_D^2)_i^{~j} \widetilde{d}^{\dagger i} \widetilde{d}_j
    + (m_U^2)_i^{~j} \widetilde{u}^{\dagger i} \widetilde{u}_j
  \nonumber\\
  &&+ (m_E^2)^i_{~j} \widetilde{e}_i \widetilde{e}^{\dagger j} 
    + (m_L^2)_i^{~j} \widetilde{l}^{\dagger i} \widetilde{l}_j
  \nonumber\\
  &&+ \Delta_1^2 h_1^\dagger h_1
    + \Delta_2^2 h_2^\dagger h_2
    - \left( B\mu h_1 h_2 + \text{H.c.}\right)
  \nonumber\\
  &&+ \left(   A_D^{ij} \widetilde{d}_i \widetilde{q}_j h_1
             + A_U^{ij} \widetilde{u}_i \widetilde{q}_j h_2
             + A_L^{ij} \widetilde{e}_i \widetilde{l}_j h_1
             + \text{H.c.}\right)
  \nonumber\\
  &&+\frac{M_1}{2} \overline{\widetilde{B}}\widetilde{B}
    +\frac{M_2}{2} \overline{\widetilde{W}}\widetilde{W}
    +\frac{M_3}{2} \overline{\widetilde{g}}\widetilde{g}~,
  \label{soft}
\end{eqnarray}
where $\widetilde{q}_i$, $\widetilde{u}_i$, $\widetilde{d}_i$,
$\widetilde{l}_i$, $\widetilde{e}_i$, $h_1$, and $h_2$  
are the corresponding scalar components of the chiral superfields,
and $\widetilde{B}$, $\widetilde{W}$, and $\widetilde{g}$ 
denote $\text{U(1)}_{\text{Y}}$, $\text{SU(2)}_{\text{L}}$, and
$\text{SU(3)}_{\text{C}}$ gauge fermions, respectively.

\subsection{Flavor Structure of the Soft Breaking Terms}
Although the Yukawa couplings are the only source
of the flavor mixing in the SM,
the mass terms and the trilinear scalar coupling terms ($A$-terms) 
of squarks and sleptons in Eq.~(\ref{soft})
may induce additional flavor mixings in the MSSM.

The Yukawa couplings $f$'s in Eq.~(\ref{MSSMsuperpotential}) are
constrained to reproduce the known quark and lepton masses and 
the CKM matrix. On the other hand, the soft breaking terms, 
their flavor structures in particular, are rather unconstrained 
at first sight, apart from the naturalness condition that
they should be within the TeV scale.
As is well-known, however, unless some specific structure
is assumed in the soft breaking terms, the sfermion-exchanging
contributions to flavor changing neutral current (FCNC) processes
such as the \kk\  mixing and the $\mu\rightarrow e\gamma$ decay 
are too large to satisfy the experimental limits, if the squark and 
slepton masses are below a few $\text{TeV}$. 
There are several ways to avoid this problem.

One is to assume a SUSY breaking (and its mediation)
mechanism in which the universality of the soft breaking terms are ensured.
In other words, mass degeneracy for the sfermions with
the same electric charge and chirality, and proportionality of
the $A$-terms to the Yukawa coupling constants are required
to suppress FCNC processes.
Phenomenology of models with the universality further depends on
the energy scale where the SUSY breaking is generated,
because the universality in the sfermion sector is vitiated due to
radiative corrections induced by the Yukawa couplings.

It is convenient to use renormalization group (RG) equations
in order to trace these radiative corrections. 
The universality in the soft breaking terms is imposed on
the boundary conditions of the RG equations 
at the energy scale of the SUSY breaking.
If the energy scale of the SUSY breaking is close to the electroweak scale, 
the RG evolution is tiny and the universality is practically maintained.
The gauge mediated SUSY breaking model\cite{Dine:1982gu}
is of this kind, and its flavor phenomenology is quite similar to that of
the SM.

If the energy scale of the SUSY breaking is far above the electroweak scale,
the RG evolution is sizable and the universality is lost. 
Though not exactly universal, flavor physics is still under control
in this kind of models in the sense that the origin of the flavor mixing 
only resides in the Yukawa couplings. The flavor mixing in the squark sector
is determined by the quark masses and the CKM matrix.
On the other hand, the flavor mixing in the slepton sector is ruled by 
the lepton
couplings in the superpotential including Majorana mass terms of
right-handed neutrinos if exist. 
The mSUGRA discussed in Sec.~{\ref{SEC:mSUGRA}} 
is a model in this class.

Embedded in a GUT, the above situation is modified if
the energy scale of the SUSY breaking is higher than the GUT scale.
Since the GUT interactions obscure the distinction between
the quark flavors and the lepton flavors,
the lepton flavor mixing in the Yukawa couplings affects the squark sector.
We shall examine an SU(5) SUSY GUT with right-handed 
neutrinos among models of this kind in Sec.~\ref{SEC:SUSYGUTnuR}.

Another way to suppress the FCNC processes is to rely 
on a flavor (or horizontal) symmetry. 
It is obvious that the flavor symmetry, whatever it is, 
should be broken because the Yukawa couplings 
have no such symmetry.
The symmetry breaking must be taken so that the observed 
quark and lepton masses and their mixings are reproduced. 
Even though this constraint is imposed, there are several choices
for the flavor symmetry and its breaking pattern.
The flavor phenomenology heavily depends on them. 
In Sec.~\ref{SEC:U2}, we shall consider a model with 
U(2) flavor symmetry among the possibilities.

\subsection{\label{SEC:mSUGRA}The Minimal Supergravity Model}
The mSUGRA consists of the observable sector, \emph{i.e.}
the MSSM, and a hidden sector. 
These two sectors are only interconnected by the gravitation.
The SUSY is assumed to be spontaneously broken 
in the hidden sector, and the soft breaking terms in Eq.~(\ref{soft})
are induced through the gravitational interaction in the following
manner:
\begin{eqnarray}
  (m_Q^2)^i_{~j} &=&
  (m_E^2)^i_{~j} ~=~ m_0^2\ \delta^i_{~j} ~,
\nonumber\\
  (m_D^2)_i^{~j} &=&
  (m_U^2)_i^{~j} ~=~
  (m_L^2)_i^{~j} ~=~ m_0^2\ \delta_i^{~j} ~,
\nonumber\\
  \Delta_1^2 &=& \Delta_2^2 ~=~ m_0^2 ~,
\nonumber\\
  A_D^{ij} &=& m_0 A_0 f_{D}^{ij}~, ~~
  A_U^{ij} ~=~ m_0 A_0 f_{U}^{ij}~, ~~
  A_L^{ij} ~=~ m_0 A_0 f_{L}^{ij}~,
  \nonumber\\
  M_1 &=& M_2 ~=~ M_3 ~=~ M_{1/2} ~,
\label{eq:boundaryconditions}
\end{eqnarray}
where we assume the GUT relation among the gaugino masses.
The above relations are applied at the energy scale where
the soft breaking terms are induced by the gravitational
interaction. We identify this scale with the GUT scale
($M_X\simeq 2\times 10^{16}\text{GeV}$) for simplicity.

The soft breaking terms at the electroweak scale are
determined by solving RG equations with the initial
conditions defined in Eq.~(\ref{eq:boundaryconditions}).

\subsection{\label{SEC:SUSYGUTnuR}
            The SU(5) SUSY GUT with Right-handed Neutrinos}
The measurements of the three gauge coupling constants at
LEP, SLC, and other experiments support the idea of
the supersymmetric grand unification. Furthermore, there is
clear evidence of neutrino oscillations in the atmospheric
\cite{Fukuda:1998mi}
and the solar\cite{Fukuda:2001nj} neutrino experiments.
Guided by these experimental results, SU(5) SUSY GUT
with right-handed neutrino has been studied. In particular,
the relationship between quark flavor signals and the neutrino
oscillations has been investigated 
in Ref.~\cite{Baek:2001sj,Moroi:2000,Baek:2001kh}.
A large flavor mixing in the neutrino sector
can include a squark mixing in the right-handed down-type squark sector.
In Ref.~\cite{Baek:2001sj}, the quark flavor signals are
studied for various neutrino oscillation scenarios.
In Ref.~\cite{Moroi:2000}, effects of CP violating phases in
the GUT Yukawa coupling constants
are taken into account.
It is shown in these papers that large contributions to
$\ek$ and the $\meg$ decay can arise from the new source of flavor mixing
in the neutrino sector. These analyses are extended to the
case of a GUT model with realistic fermion mass matrices in 
Ref.~\cite{Baek:2001kh}. Here we follow the analysis of 
Ref.~\cite{Baek:2001kh}
and we give a brief description of the model.

The Yukawa couplings and Majorana masses of the right-handed 
neutrinos in the model are described by the 
following superpotential:
\begin{eqnarray}
\mathcal{W}_{\text{SU(5)}\nu_R} &=&
 \frac{1}{8}\epsilon_{abcde}(\lambda_U)^{ij}(T_i)^{ab}(T_j)^{cd}H^e
 +(\lambda_D)^{ij}(\overline{F}_i)_a(T_j)^{ab
}\overline{H}_b
\nonumber\\&&
 +(\lambda_N)^{ij}\overline{N}_i(\overline{F}_j)_a H^a
 +\frac{1}{2}(M_N)^{ij}\overline{N}_i\overline{N}_j,
\label{eq:SU5RN}
\end{eqnarray}
where $i$ and $j$ are generation indices, while $a,\,b,\,c,\,d,$ and 
$e$ are SU(5) indices.
$\epsilon_{abcde}$ denotes the totally antisymmetric tensor of 
the SU(5). $T_i$, $\overline{F}_i$, and $\overline{N}_i$ are 
${\bf 10}$, ${\bf \overline{5}}$, and ${\bf 1}$ representations of
the SU(5) gauge group, respectively.
$T_i$ contains $Q_i$, $\overline{U}_i$, and $\overline{E}_i$ 
in Eq.~(\ref{MSSMsuperfields}), and
$\overline{F}_i$ includes $\overline{D}_i$ and $L_i$.
$H$ and $\overline{H}$ are Higgs superfields in ${\bf 5}$ and
${\bf \overline{5}}$ representations, respectively.
$H$ consists of $H_C({\bf 3,1},-\frac{1}{3})$ and $H_2$, and
$\overline{H}$ does $\overline{H}_C({\bf \overline{3},1},\frac{1}{3})$
and $H_1$. 
$(\lambda_U)^{ij}$, $(\lambda_D)^{ij}$, and $(\lambda_N)^{ij}$
are the Yukawa coupling matrices, and $(M_N)^{ij}$ is
the Majorana mass matrix.
In addition to the above superpotential, we also need a superpotential 
for Higgs superfields, $\mathcal{W}_H(H,\overline{H},\Sigma)$, where
$\Sigma^a_{~b}$ is a {\bf 24} representation of the SU(5) group.
It is assumed to develop a vacuum expectation value (VEV) as
$\langle \Sigma^a_{~b} \rangle=
\text{diag}(\frac{1}{3},\frac{1}{3},\frac{1}{3},
-\frac{1}{2},-\frac{1}{2})v_G$ at the GUT scale
and breaks the SU(5) symmetry to $\text{SU(3)}_{\text{C}}\times
\text{SU(2)}_{\text{L}}\times\text{U(1)}_{\text{Y}}$.

The supermultiplets whose masses are of order of the GUT scale
such as $H_C$ and $\overline{H}_C$ can be integrated out
below the GUT scale. Then, the effective theory below the GUT
scale is the MSSM with the right-handed neutrino supermultiplets,
and its superpotential is given as
\begin{equation}
\mathcal{W}_{\text{MSSM}\nu_R}=\mathcal{W}_{\text{MSSM}}
 + (f_{N})^{ij}\overline{N}_i L_jH_2
 + \frac{1}{2}(M_N)^{ij}\overline{N}_i\overline{N}_j,
\label{eq:MSSMRN}
\end{equation}
where the Yukawa coupling matrices are related to those in 
Eq.~(\ref{eq:SU5RN}) as $(f_U)^{ij}=(\lambda_U)^{ij}$,
$(f_D)^{ij}=(f_E^T)^{ij}=(\lambda_D)^{ij}$, and
$(f_{N})^{ij}=(\lambda_N)^{ij}$ in the leading order
approximation.

In the energy region lower than the Majorana mass scale ($\equiv M_R$),
the singlet supermultiplets are integrated out, and the
resulting superpotential is the sum of $\mathcal{W}_{\text{MSSM}}$
in Eq.~(\ref{MSSMsuperpotential}) and the following
higher dimensional term:
\begin{equation}
\Delta\mathcal{W}_{\nu} = -\frac{1}{2}(K_{\nu})^{ij}(L_i H_2)(L_j H_2),
\qquad K_{\nu} = (f_N^T)^{ik}(\frac{1}{M_N})_{kl}(f_N)^{lj}.
\label{eq:seesaw1}
\end{equation}
This term yields the neutrino masses below the electroweak scale as
\begin{equation}
(m_{\nu})^{ij} = (K_{\nu})^{ij}\langle h_2 \rangle^2.
\label{eq:seesaw2}
\end{equation}
The above neutrino mass matrix is related to the observable
neutrino mass eigenvalues  and the Maki-Nakagawa-Sakata (MNS) mixing 
matrix\cite{Maki:1962mu} as
\begin{equation}
(m_{\nu})^{ij} =(V_{\text{MNS}}^*)^i_{~k}
                {m_{\nu}}^k(V_{\text{MNS}}^{\dag})_k^{~j}
\label{eq:MNS}
\end{equation} 
in the basis in which the charged lepton mass matrix
is diagonal.

As is mentioned above, the superpotential in Eq.~(\ref{eq:SU5RN})
predicts that
\begin{equation}
(f_E)^{ij}=(f_D)^{ji}
\label{NSU5REL}
\end{equation}
at the GUT scale. 
It is well-known that the mass ratios of the down-type quarks to
the charged leptons in the first and the second generations 
can not be explained by this relation although it reasonably 
works for the third generation. This, however, is not a fatal
flaw of SU(5) GUTs because there are several ways to overcome
this shortcoming. For example, higher dimensional operators with 
$\Sigma^a_{~b}$ may contribute differently to the Yukawa coupling
matrices of the down-type quarks and the charged leptons.

In Ref.~\cite{Baek:2001kh}, quark FCNC processes, lepton flavor
violation processes, and the muon anomalous magnetic moment
were calculated in this model. To account for quark and lepton
mass ratios, the following higher dimensional operator was
introduced: 
\begin{equation}
\Delta\mathcal{W}_{\text{SU(5)}\nu_R}=
\frac{(\kappa_D)^{ij}}{M_P}
(\overline{F}_i)_a\Sigma^a_{~b}(T_j)^{bc}\overline{H}_c,
\end{equation}
where $M_P$ is the Planck mass ($M_P\simeq 2\times 10^{18}$ GeV).
Consequently, Eq.~(\ref{NSU5REL})
is modified to
\begin{equation}
  (f_E)^{ij}=(f_D)^{ji}-\frac{5}{6}\xi(\kappa_D)^{ji},
\end{equation}
where $\xi=v_G/M_P\simeq 0.01$. 
Taking the Majorana mass matrix proportional to the unit matrix
($(M_N)^{ij}=M_R\delta^{ij}$) for simplicity,
Baek {\it et al.}~\cite{Baek:2001kh} showed that the flavor 
mixings of the squark and slepton 
sectors were determined by the CKM matrix, the MNS matrix, and
two additional mixing matrices related to the down-type
quark and the charged lepton Yukawa coupling constants.
As long as we take the large mixing MSW solution for the solar
neutrino anomaly, the SUSY effect becomes large for
$\varepsilon_K$ and B($\mu\rightarrow e\gamma$).
In this paper we follow Ref.~\cite{Baek:2001kh} but consider the special 
case that the two additional
mixing matrices are equal to the unit matrix, because the general
features mentioned above do not change by this simplification.

In order to calculate the FCNC processes
we need to specify the soft breaking terms.
The SU(5) invariant soft breaking terms are written as
\begin{eqnarray}
  -\mathcal{L}^{\text{SU(5)}}_{\text{soft}} &=&
  (m^2_T)_i^{~j}(\widetilde{T}^{i*})_{ab}(\widetilde{T}_j)^{ab}
  +(m^2_{\overline{F}})_i^{~j}
   (\widetilde{\overline{F}}{}^{i*})^a (\widetilde{\overline{F}}_j)_a
  +(m^2_{\overline{N}})_i^{~j}
   \widetilde{\overline{N}}{}^{i*}\widetilde{\overline{N}}_j
\nonumber\\&&
  +(m^2_{H}) H^*_{~a} H^a
  +(m^2_{\overline{H}}) \overline{H}^{*a} \overline{H}_a
\nonumber\\&&
  +\left\{
      \frac{1}{8}\epsilon_{abcde}(\widetilde{\lambda}_U)^{ij}
      (\widetilde{T}_i)^{ab}(\widetilde{T}_j)^{cd} H^e
    + (\widetilde{\lambda}_D)^{ij}(\widetilde{\overline{F}}_i)_a
      (\widetilde{T}_j)^{ab} \overline{H}_b
    \right.
\nonumber\\&&\phantom{+}
    \left.
    + (\widetilde{\lambda}_N)^{ij}
      \widetilde{\overline{N}}_i(\widetilde{\overline{F}}_j)_a H^a
    + \frac{1}{2}(\widetilde{M}_N)^{ij}
      \widetilde{\overline{N}}_i\widetilde{\overline{N}}_j
      + \text{H.c.} \right\}
\nonumber\\&&
+\frac{1}{M_P}\left[(\widetilde{\kappa}_d)^{ij}
     (\widetilde{\overline{F}}^i)_a\Sigma^a_{~b}(\widetilde{T}^j)^{bc}
     \overline{H}_c+\text{H.c.}
   \right]
\nonumber\\&&
    +\frac{1}{2}M_5\overline{\widetilde{G}_5}\widetilde{G}_5,
\label{eq:SUSY-Breaking-SU5RN}
\end{eqnarray}
where $\widetilde{T}^i$, ${\widetilde{\overline{F}}}{}^i$, and
${\widetilde{\overline{N}}}{}^i$ are the scalar components of $T^i$,
$\overline{F}^i$, and $\overline{N}^i$, respectively;
$H$ and $\overline H$ stand for the corresponding scalar components
of the superfields denoted by the same symbols; and
$\widetilde{G}_5$ represents the SU(5) gaugino.
We assume that the soft breaking terms are universally generated
at the Planck scale, \emph{i.e.}
\begin{eqnarray}
  (m^2_T)_i^{~j} &=&
  (m^2_{\overline{F}})_i^{~j} =
  (m^2_{\overline{N}})_i^{~j} =
  m_0^2\delta_i^{j},\nonumber\\
  (\widetilde{\lambda})^{ij} &=& m_0 A_0 (\lambda)^{ij},\quad
  ( \lambda = \lambda_U, \lambda_D, \lambda_N ), \nonumber\\
  (\widetilde{\kappa}_D)^{ij} &=& m_0 A_0 (\kappa_D)^{ij},\nonumber\\
  M_5 &=& M_{{1/2}}.
\end{eqnarray}
These equations serve as a set of boundary conditions of
RG equations at the Planck scale.

We solve the RG equations of the SU(5) SUSY GUT from the Planck
scale to the GUT scale, then those of MSSM with right-handed neutrinos
between the GUT scale and $M_R$.
Finally, the squark and slepton mass matrices are obtained by the RG 
equations of the MSSM below $M_R$.

\subsection{\label{SEC:U2}A Model with U(2) Flavor Symmetry}
It is possible that the family structure of the quarks and the leptons is
explained by some flavor symmetry. 
Although U(3) is a natural candidate of the flavor symmetry, it is
badly broken by the top Yukawa coupling.
It is therefore legitimate to choose a U(2) symmetric model in order to
study the flavor problem in the MSSM.

In this framework, the quark and lepton supermultiplets in 
the first and the second generations
transform as doublets under the U(2) flavor symmetry. Each of these
doublets carries a positive unit charge of the U(1) subgroup. 
The quark and lepton supermultiplets in the third generation 
and the Higgs supermultiplets are totally singlet under the U(2).
In addition to the ordinary matter fields, we introduce the following
superfields: a doublet $\Phi^i(-1)$, a symmetric tensor $S^{ij}(-2)$, 
and an antisymmetric tensor $A^{ij}(-2)$, where $i$ and $j$ run from 1 to 2,
and the numbers in the parentheses represent the U(1) 
charges\cite{Barbieri:1997ww}.

The U(2) invariant superpotential relevant to the quark
Yukawa couplings is given as follows:
\begin{eqnarray}
\mathcal{W}_{\text{U(2)}} &=& y_U\left({\overline U}_3 Q_3 H_2
                    +\frac{b_U}{M_F}\Phi^i{\overline U}_i Q_3 H_2
                    +\frac{c_U}{M_F}{\overline U}_3\Phi^i Q_i H_2
                    \right.\nonumber\\
                &&  \left.+\frac{d_U}{M_F}S^{ij}{\overline U}_i Q_j H_2
                    +\frac{a_U}{M_F}A^{ij}{\overline U}_i Q_j H_2\right)
                    \label{U2superpotential}\\
                &&  +y_D\left({\overline D}_3 Q_3 H_1
                    +\frac{b_D}{M_F}\Phi^i{\overline D}_i Q_3 H_1
                    +\frac{c_D}{M_F}{\overline D}_3\Phi^i Q_i H_1
                    \right.\nonumber\\
                &&  \left.+\frac{d_D}{M_F}S^{ij}{\overline D}_i Q_j H_1
                    +\frac{a_D}{M_F}A^{ij}{\overline D}_i Q_j H_1\right)
                    \nonumber,
\end{eqnarray}
where $M_F$ is the scale of the flavor symmetry, and 
$y_Q$, $a_Q$, $b_Q$, $c_Q$, and $d_Q$ ($Q=U,D$) are dimensionless
coupling constants. In Eq.~(\ref{U2superpotential}), we neglected
dimension five and higher dimensional operators in the superpotential.
Absolute values of the above dimensionless coupling constants
except for $y_D$ are supposed to be of $O(1)$.

The successful breaking pattern of the U(2) symmetry is that
\begin{equation}
\text{U(2)}\longrightarrow\text{U(1)}\longrightarrow\openone
(\text{no symmetry}),
\label{U2breaking}
\end{equation}     
where the first breaking is induced by VEV's of $\Phi^i$ and $S^{ij}$,
and a VEV of $A^{ij}$ brings about the second one. The VEV's are
given as
\begin{equation}
\frac{\langle\Phi^i\rangle}{M_F}=\delta^{i2}\,\epsilon_\Phi,\;
\frac{\langle S^{ij}\rangle}{M_F}=\delta^{i2}\,\delta^{j2}\epsilon_S,\;
\frac{\langle A^{ij}\rangle}{M_F}=\epsilon^{ij}\,\epsilon',
\end{equation}
where $\epsilon_\Phi$ and $\epsilon'$ are taken to be real without
loss of generality. Note that $\langle S^{ij}\rangle$ is chosen 
so that it leaves a U(1) unbroken.
Hierarchical relations among the VEV's that 
$\epsilon'\ll\epsilon_\Phi\sim|\epsilon_S|\ll 1$ are assumed
in order to reproduce the quark masses and the quark mixing angles.

With the above VEV's, we obtain the following quark Yukawa couplings:
\begin{equation}
(f_Q^{ij})=y_Q \left(\begin{array}{ccc}
                      0               & a_Q\,\epsilon' & 0          \\
                      -a_Q\,\epsilon' & d_Q\,\epsilon  & b_Q\,\epsilon\\
                      0               & c_Q\,\epsilon  & 1          \\
                     \end{array}\right),\quad Q=U,D,
\label{U2Yukawa}
\end{equation}
where we use $\epsilon\equiv\epsilon_\Phi=\epsilon_S$,
which is valid providing appropriate redefinitions of the coupling 
constants in Eq.~(\ref{U2superpotential}).
Eq.~(\ref{U2Yukawa}) is applied at the GUT scale where we assume 
that the symmetry breaking sequence in Eq.~(\ref{U2breaking}) occurs.
The parameters in Eq.~(\ref{U2Yukawa}) are determined so that
the known quark masses and mixing are reproduced taking the RG
evolution into account.

The U(2) symmetry constrains the soft breaking terms as well as the
supersymmetric terms. The U(2) invariant soft breaking terms
relevant to squark masses are
\begin{eqnarray}
-\mathcal{L}_m = \sum_{f=q,u,d} m_0^{f2}
               & & \left[
                   \widetilde f^{*i}\widetilde f_i
                   +a_f^3 \widetilde f^{*3} \widetilde f_3
                   +\frac{a_f^\phi}{M_F}
                    \widetilde f^{*3}\phi^i\widetilde f_i
                   +\frac{a_f^{\phi *}}{M_F}
                    \phi_i^*\widetilde f^{i*}\widetilde f_3
                   \right.\nonumber\\
               & & \left.
                   +\frac{a_f^{\phi\phi}}{M_F}
                    \phi_i^*\widetilde f^{i*}\phi^j\widetilde f_j
                   +\frac{a_f^{SS}}{M_F}
                    S_{ik}^*\widetilde f^{k*}S^{ij}\widetilde f_j
                   \right],
\label{U2smass}
\end{eqnarray}
where $a_f$'s are dimensionless coupling constants of $O(1)$,
and shown are the terms that yield squark masses of $O(\epsilon^2)$
or larger when the flavor symmetry breaking takes place. 
The squark mass matrices stemming from Eq.~(\ref{U2smass})
are parameterized as:
\begin{equation}
m_X^2=m_0^{X2}\left(\begin{array}{ccc}
                    1 & 0 & 0 \\
                    0 & 1+r_{22}^X\epsilon^2 & r_{23}^{X}\epsilon \\
                    0 & r_{23}^{X*}\epsilon & r_{33}^{X}
		    \end{array}
              \right),\quad X=Q,U,D,
\label{U2smassmatrix}
\end{equation}
where $r^X$'s are constant parameters of $O(1)$. 

As for the $A$-terms, it turns out that they have the same hierarchical
structure as the Yukawa couplings in Eq.~(\ref{U2Yukawa}):
\begin{equation}
(A_Q^{ij})=A_Q^0
 \left(\begin{array}{ccc}
        0               & \tilde a_Q\,\epsilon' & 0          \\
        -\tilde a_Q\,\epsilon' & \tilde d_Q\,\epsilon  & \tilde b_Q\,\epsilon\\
        0               & \tilde c_Q\,\epsilon  & 1          \\
       \end{array}\right),\quad Q=U,D.
\label{U2Aterm}
\end{equation}
In general, though being of $O(1)$, $\tilde a_Q$, $\tilde b_Q$, $\tilde c_Q$, 
and $\tilde d_Q$ take different values from the corresponding 
parameters in Eq.~(\ref{U2Yukawa}). Therefore, we expect no exact universality
of the $A$-terms in this model.

The soft SUSY breaking terms at the electroweak scale are given by
solving the  RG equations of the MSSM with the
boundary conditions in Eq.~(\ref{U2smassmatrix}) and Eq.~(\ref{U2Aterm})
at the GUT scale.

%%%%%%%%%%%%%%%%%%%%%%%%%%%%%%%%%%%%%%%%%%%%%%
% Section 3 of the paper by T.Goto, Y.Okada, %
% Y.Shimizu, T.Shindou and M.Tanaka.         %
% Written by T.G.                            %
%%%%%%%%%%%%%%%%%%%%%%%%%%%%%%%%%%%%%%%%%%%%%%
\section{Observables\label{OBSERVABLES}}

The observables considered in the following are the CP violation 
parameter $\varepsilon_K$ in the $K^0-\bar{K}^0$ mixing,
$B_d-\bar{B}_d$ and $B_s-\bar{B}_s$ mass splittings $\Delta m_{B_d}$
and $\Delta m_{B_s}$, respectively and CP asymmetries in
various $B$ decay modes.

The \bdbd, \bsbs, and \kk\ mixings are described by the effective
Lagrangian of the following form:
\begin{eqnarray}
{\cal L} &=&
  C_{LL}      
  (\bar{q}_L^{\alpha} \gamma^\mu Q_{L\alpha})
  (\bar{q}_L^{\beta } \gamma_\mu Q_{L\beta })
+ C_{RR}
  (\bar{q}_R^{\alpha} \gamma^\mu Q_{R\alpha})
  (\bar{q}_R^{\beta } \gamma_\mu Q_{R\beta })
\nonumber\\&&
+ C_{LR}^{(1)}
  (\bar{q}_R^{\alpha} Q_{L\alpha})
  (\bar{q}_L^{\beta } Q_{R\beta })
+ C_{LR}^{(2)}
  (\bar{q}_R^{\alpha} Q_{L\beta })
  (\bar{q}_L^{\beta } Q_{R\alpha})
\nonumber\\&&
+ \tilde{C}_{LL}^{(1)}
  (\bar{q}_R^{\alpha} Q_{L\alpha})
  (\bar{q}_R^{\beta } Q_{L\beta })
+ \tilde{C}_{LL}^{(2)}
  (\bar{q}_R^{\alpha} Q_{L\beta })
  (\bar{q}_R^{\beta } Q_{L\alpha})
\nonumber\\&&
+ \tilde{C}_{RR}^{(1)}
  (\bar{q}_L^{\alpha} Q_{R\alpha})
  (\bar{q}_L^{\beta } Q_{R\beta })
+ \tilde{C}_{RR}^{(2)}
  (\bar{q}_L^{\alpha} Q_{R\beta })
  (\bar{q}_L^{\beta } Q_{R\alpha}),
\label{eq:lag}
\end{eqnarray}
where $(q,\,Q)=(d,\,b)$, $(s,\,b)$ and $(d,\,s)$ for the \bdbd, \bsbs,
and \kk\ mixings, respectively.
The suffices $\alpha$ and $\beta$ are color indices.
The Wilson coefficients $C$'s are obtained by calculating box diagrams.
See Ref.~\cite{Baek:2001kh} for explicit formulas of the coefficients.
The mixing matrix elements $M_{12}(B_d)$, $M_{12}(B_s)$, and $M_{12}(K)$ 
are given as
\begin{equation}
  M_{12}(P) = -\frac{1}{2m_P}
  \langle P | {\cal L} | \bar{P} \rangle,
\end{equation}
where $P=B_d,\,B_s,\,K^0$.

In the SM, the flavor changes only occur in the
interaction with the left-handed quarks, so that $M_{12}$ is dominated
by the $C_{LL}$ term for all the three mesons.
The situation is the same in the mSUGRA, since the
flavor mixing in the squark sector is induced by the running effect and
hence takes place only in the left-handed squark sector.
In the SU(5) SUSY GUT with right-handed neutrinos and
the U(2) model, on the other hand, there are sources of squark flavor
mixing other than the CKM matrix.
In the SU(5) SUSY GUT with right-handed neutrinos, flavor mixing
in the right-handed down-type squark sector is induced due to the Yukawa
coupling matrix of the neutrinos through the running between the GUT and
the Planck scales.
In the U(2) model, the squark mass matrices contain more free parameters.
Consequently, flavor mixing is possible in 
both the left-handed and the right-handed squark sectors
and the mixing matrices can be different from the CKM matrix.
In the latter two models, all the Wilson coefficients in Eq.~(\ref{eq:lag}) 
are relevant.

We parameterize the matrix elements of the operators in Eq.~(\ref{eq:lag}) as
\begin{subequations}
\begin{eqnarray}
  \langle K^0 |
  (\bar{d}_L^{\alpha} \gamma^\mu s_{L\alpha})
  (\bar{d}_L^{\beta } \gamma_\mu s_{L\beta })
  | \bar{K}^0 \rangle
  &=&
  \frac{2}{3}m_K^2 f_K^2 B_K,
\\
  \langle K^0 |
  (\bar{d}_R^{\alpha} s_{L\alpha})
  (\bar{d}_L^{\beta } s_{R\beta })
  | \bar{K}^0 \rangle
  &=&
  \frac{1}{2} \left( \frac{m_K}{m_s+m_d} \right)^2
  m_K^2 f_K^2 B_K^{LR(1)},
\\
  \langle K^0 |
  (\bar{d}_R^{\alpha} s_{L\beta })
  (\bar{d}_L^{\beta } s_{R\alpha})
  | \bar{K}^0 \rangle
  &=&
  \frac{1}{6} \left( \frac{m_K}{m_s+m_d} \right)^2
  m_K^2 f_K^2 B_K^{LR(2)},
\\
  \langle K^0 |
  (\bar{d}_L^{\alpha} s_{R\alpha})
  (\bar{d}_L^{\beta } s_{R\beta })
  | \bar{K}^0 \rangle
  &=&
  -\frac{5}{12} \left( \frac{m_K}{m_s+m_d} \right)^2
  m_K^2 f_K^2 \tilde{B}_K^{RR(1)},
\\
  \langle K^0 |
  (\bar{d}_L^{\alpha} s_{R\beta })
  (\bar{d}_L^{\beta } s_{R\alpha})
  | \bar{K}^0 \rangle
  &=&
  \frac{1}{12} \left( \frac{m_K}{m_s+m_d} \right)^2
  m_K^2 f_K^2 \tilde{B}_K^{RR(2)},
\end{eqnarray}
\label{eq kk matrix element}
\end{subequations}
where $B_K$, $B_K^{LR(1,2)}$, and $\tilde{B}_K^{RR(1,2)}$ are bag
parameters of $O(1)$, which have been calculated by the lattice QCD 
method\cite{Yamada:2000kt}.
It can be seen that the matrix elements of the scalar operators are
enhanced by a factor $\sim(m_K/m_s)^2$ for the \kk\ mixing.
For \bb\ mixing cases, the corresponding factor is $\sim(m_B/m_b)^2$ so
that the enhancement is less significant. In the SU(5) SUSY GUT with 
right-handed neutrinos and the U(2) model,
the $C_{LR}$ and/or $\tilde{C}_{RR}$ terms
can significantly contribute to $M_{12}(K)$ because of this enhancement in
the matrix elements.

We can express 
$\varepsilon_K$, $\dmbd$, and $\dmbs$ in terms of $M_{12}$ as
\begin{eqnarray}
\ek &=& \frac{ \e^{i\pi/4}\im M_{12}(K) }{ \sqrt{2}\Delta m_K },
\\
\dmbd &=& 2\left| M_{12}(B_d) \right|,
\\
\dmbs &=& 2\left| M_{12}(B_s) \right|.
\end{eqnarray}
The CP asymmetry in $B\to J/\psi\,K_S$,
\acp, is defined as
\begin{eqnarray}
 \frac{
   \Gamma( B_d(t)\to J/\psi\,K_S ) - \Gamma( \bar{B}_d(t)\to J/\psi\,K_S )
   }{
   \Gamma( B_d(t)\to J/\psi\,K_S ) + \Gamma( \bar{B}_d(t)\to J/\psi\,K_S )
   } 
 =
 - \acp\,
 \sin\dmbd t.
\end{eqnarray}
This asymmetry is given by the phase of $M_{12}(B_d)$ as
\begin{equation}
\acp = \sin\phi_M\;,
\end{equation}
where $\phi_M$ is defined as
$\e^{i\phi_M} = M_{12}(B_d)/| M_{12}(B_d) |$.
In the present analysis we assume that tree-level diagram dominates
the $B_d(\bar{B}_d)\to J/\psi\,K_S$ decay so that no new phase appears in
the decay amplitude\footnote{
In the U(2) model, in particular, there might be sizable 
contributions to the decay amplitudes with new CP phases. In such a
case a direct CP asymmetry may be observed.}.
Experimentally, $\sin\phi_M 
$ can be determined by combining decay
modes with the $b \rightarrow c\overline{c}s$ transition such as 
$B_d \to J/\psi\,K_S$, $B_d \to J/\psi\,K_L$,
and $B_d \to \psi'\,K_S$.

\begin{figure}
\includegraphics{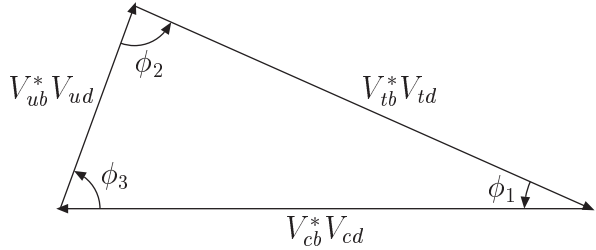}
\caption{
  The unitarity triangle.
}
\label{fig:ut}
\end{figure}

In order to constrain new physics from the consistency check
on the closure of the unitarity triangle, depicted in FIG.~\ref{fig:ut},
it is important to measure the angles other than $\phi_M$.
There are several theoretically clean ways to determine these angles.
In the SM, $2\phi_1$ is given by $\phi_M$, and
$\sin 2\phi_2$ is obtained from the isospin analysis
of $B\to\pi\pi$\cite{Gronau:1990ka} and the time-dependent Dalitz
analysis of $B\to\rho\pi$\cite{Lipkin:1991st}.
$B\to D^{(*)}K^{(*)}$ modes provide us with
the angle $\phi_3$\cite{Gronau:1991dp}. 
$B\to D^{(*)}\pi$\cite{Dunietz:1988bv} and
$B\to D^{*}\rho$\cite{London:2000zi} with the angular analysis
give us information on $\sin(2\phi_1+\phi_3)$.

If we consider effects of new physics, these measurements can be
interpreted as constraints on the phases of the \bdbd\ mixing
and the decay amplitudes.
When we neglect the new phases in the decay amplitudes,
$B\to\pi\pi,\,\rho\pi$, $B\to D^{(*)}K^{(*)}$, and
$B\to D^{(*)}\pi,\,D^{*}\rho$ provide us with $\sin(\phi_M+2\phi_3)$,
$\phi_3$, and $\sin(\phi_M+\phi_3)$, respectively, where $\phi_3$ is the
weak phase of the $b\to u$ transition amplitude in the standard phase
convention given in APPENDIX~\ref{app:CKM}\footnote{
This approximation is valid for the three models under consideration,
at least for $B\to D^{(*)}K^{(*)},\,D^{(*)}\pi,\,D^{*}\rho$.
New phases could be important for the decay amplitudes of
$B\to\pi\pi,\,\rho\pi$. Even in such a case,
we could obtain information about new phases by measuring
CP asymmetries of the various modes listed above.}.
In the following analysis,
we assumed that $\phi_M$ is determined from the $B\to J/\psi\,K_S$ mode
and related modes, and
we study impacts of the $\phi_3$ measurement on new physics search.

%%%%%%%%%%%%%%%%%%%%%%%%%%%%%%%%%%%%%%%%%%%%%%
% Section 4 of the paper by T.Goto, Y.Okada, %
% Y.Shimizu, T.Shindou and M.Tanaka.         %
% Written by T.G.                            %
%%%%%%%%%%%%%%%%%%%%%%%%%%%%%%%%%%%%%%%%%%%%%%
\section{Numerical ANALYSIS\label{NR}}

\subsection{
Parameters in the minimal supergravity model
\label{sec parameters in mSUGRA}
}

In our calculation, the masses and the mixing matrices in the quark and
lepton sectors are treated as input parameters which determine the Yukawa
coupling matrices.

The CKM matrix elements $V_{us}$, $V_{cb}$, and $|V_{ub}|$
are determined in experiments independently of new physics
contributions because these are extracted from tree-level processes. 
We fix $V_{us}$ and $V_{cb}$ in the following calculations as 
$V_{us}=0.2196$ and $V_{cb}=0.04$, and
vary $|V_{ub}|$ within a range $|V_{ub}/V_{cb}|=0.09\pm0.01$.
Although the current error of $|V_{ub}|$ is estimated to be larger than
this value, we expect theoretical and experimental improvements 
in near future. 
We vary the CP violating phase, $\phi_{3}$, within $\pm 180^\circ$ 
because it is not constrained by the 
tree-level processes independently of new physics contributions.
For the quark masses, we take 
$m_t^{\text{pole}}=175\;\text{GeV}$, $m_b^{\text{pole}}=4.8\;\text{GeV}$,
$m_c^{\text{pole}}=1.4\;\text{GeV}$, and 
$m_s^{\overline{\text{MS}}}(2\;\text{GeV})=120\;\text{MeV}$.

As for the SUSY parameters, we assume $M_{1/2},\,A_0,$ and $\mu$ are
real parameters in order to avoid too large SUSY contributions to the
electric dipole moments of the neutron and the electron. We vary these 
parameters within the ranges $0<m_0<3\; \text{TeV}$, $0<M_{1/2}<1\;\text{TeV}$,
and $-5<A_0<+5$ at the GUT scale. Both signs of $\mu$ are considered.
We take the ratio of two VEV's 
$\tan\beta=\langle h_2\rangle/\langle h_1\rangle =20$ for
most of our analysis and comment on other cases.

\subsection{
	Parameters in the SU(5) SUSY GUT with right-handed neutrinos
}

In the SU(5) SUSY GUT with right-handed neutrinos, 
we need to specify the parameters in the neutrino
sector in addition to the quark Yukawa coupling constants given
in Sec.~\ref{sec parameters in mSUGRA}.
We take the neutrino masses as
$m_{\nu_3}^2 - m_{\nu_2}^2 = 2.4\times10^{-3}$ eV$^2$,
$m_{\nu_2}^2 - m_{\nu_1}^2 = 4.2\times10^{-5}$ eV$^2$, and 
$m_{\nu_1}\sim0$,
and the MNS matrix as
\begin{eqnarray}
V_{\rm MNS} &=&
\left(
\begin{array}{ccc}
  c_{\rm sol} c_{13}
& s_{\rm sol} c_{13}
& s_{13}
\\
- s_{\rm sol} c_{\rm atm}
- c_{\rm sol} s_{\rm atm} s_{13}
&
  c_{\rm sol} c_{\rm atm}
- s_{\rm sol} s_{\rm atm} s_{13}
&
  s_{\rm atm} c_{13}
\\
  s_{\rm sol} s_{\rm atm}
- c_{\rm sol} c_{\rm atm} s_{13}
&
- c_{\rm sol} s_{\rm atm}
- s_{\rm sol} c_{\rm atm} s_{13}
&
  c_{\rm atm} c_{13}
\end{array}
\right),
\end{eqnarray}
($c_i=\cos\theta_i$, $s_i=\sin\theta_i$)
with
$\sin^22\theta_{\rm atm} = 1$,
$\sin^22\theta_{\rm sol} = 0.655$, and
$\sin^22\theta_{13} = 0.015$.
These mass differences and mixing angles are consistent with the 
solar and atmospheric neutrino oscillations.
The value of $\sin^22\theta_{13}$ is constrained by reactor experiments
\cite{neutrinoreactor},
and the above value is take as an illustration.

In addition, we assume that the mass matrix of the right-handed neutrino
in Eq.~(\ref{eq:seesaw2})
is proportional to the unit matrix, and we take $M_R=4\times10^{13}$ GeV.
Complex phases in the MNS matrix and the right-handed neutrino mass
matrix are neglected.

The soft SUSY breaking parameters in this
model are assumed to be universal at 
the Planck scale, and the
running effect between the Planck and the GUT scales is taken into
account.
We scan the same ranges for $m_0$, $M_{1/2}$ and $A_0$ as those in the
mSUGRA case.

\subsection{
	Parameters in the U(2) model
}

In the U(2) model, we take the symmetry breaking
parameters $\epsilon$ and $\epsilon'$ as $\epsilon=0.04$ and
$\epsilon'=0.008$,
and the other parameters in the quark Yukawa coupling matrices in 
Eq.~(\ref{U2Yukawa}) are determined so that the CKM matrix and 
the quark masses given in Sec.~\ref{sec parameters in mSUGRA} are reproduced.
Note that the texture of the Yukawa coupling matrices in Eq.~(\ref{U2Yukawa}) 
predicts the following relation among quark masses and CKM matrix
elements:
\begin{eqnarray}
  \frac{ m_u }{ m_c } &=&
  \left| \frac{V_{ub}}{V_{cb}} \right|^2
  \left( 1 + \left| \frac{V_{ub}}{V_{cb}} \right|^2 \right),
\\
  \frac{ m_d }{ m_s } &=&
  \left| \frac{V_{td}}{V_{ts}} \right|^2
  \left( 1 + \left| \frac{V_{td}}{V_{ts}} \right|^2 \right).
\end{eqnarray}
In the numerical calculation we adjust $m_u$ and $m_d$ to satisfy these
relations.

There are many free parameters in the SUSY breaking sector as shown in 
Eq.~(\ref{U2smassmatrix}) and Eq.~(\ref{U2Aterm}). In order to reduce
the number of free parameters for numerical calculations, we assume that
\begin{eqnarray}
  m_0^{Q2} &=& m_0^{U2} = m_0^{D2} \equiv m_0^2,
\\
  r_{ij}^Q &=& r_{ij}^U = r_{ij}^D \equiv r_{ij},
\quad (ij)=(22),(23),(33).
\end{eqnarray}
We vary these parameters within the ranges $0<m_0<3$ TeV,
$-1<r_{22}<+1$, $0<r_{33}<4$, $|r_{23}|<4$ and 
$-180^\circ<\arg r_{23}<180^\circ$.
The boundary conditions for the $A$ parameters and the slepton mass matrices 
are assumed to be the same as the mSUGRA case
to simplify the numerical analysis.
We think that the above assumptions on the soft
breaking terms are sufficient for our purpose of 
comparing new physics effects related to the $B^0-\bar{B}^0$ and
the $K^0-\bar{K}^0$ mixings in the three models.

\subsection{Experimental constraints}

In order to obtain allowed parameter regions, we impose the following
experimental constraints:
\begin{itemize}
\item Lower limits on the masses of SUSY particles and the Higgs bosons
  given by the direct search in collider experiments\cite{sparticle:higgs}.
\item Branching ratio of the \bsg\ decay:
$2\times10^{-4}<\br(\bsg)<4.5\times10^{-4}$\cite{bsgamma}.
\item Branching ratio of the $\mu\to e\,\gamma$ decay for the SUSY GUT case:
$\br(\meg)<1.2\times10^{-11}$\cite{Brooks:1999pu}.
\item Measured values of $\varepsilon_K$ and 
$\Delta m_{B_d}$\cite{Groom:2000in}, and the
  lower bound of $\Delta m_{B_s}$\cite{dmbs}.
\item CP asymmetry in the $B\to J/\psi\,K_S$ decay and 
      related modes observed in the $B$ factory
  experiments\cite{Abashian:2001pa,Aubert:2001sp}.
\end{itemize}
Although the values of \ek\ and \dmbd\ are
precisely measured in experiments, there are theoretical uncertainties
in the evaluation of the matrix elements for $\Delta S(B) = 2$ operators.
In order to take these theoretical uncertainties into account, we
calculate \ek\ and \dmbd\ with bag parameters and $f_{B_{d,s}}$
in Table~\ref{tab f and bag} 
and allow parameter sets if the calculated values of \ek\ and \dmbd\
lie within the ranges
\begin{eqnarray}
\ek &=& \e^{i\pi/4} (2.28\times 10^{-3})\times(1 \pm 0.15),
\\
\dmbd &=& 0.479\,{\rm ps}^{-1}\times(1 \pm 0.2)^2.
\label{eq:dmbdrange}
\end{eqnarray}
For \dmbs, we impose a constraint on the ratio to \dmbd\ as
$\dmbs/\dmbd>27$ since a large portion of the theoretical uncertainties
is expected to cancel by taking the ratio. For the CP asymmetry,
we use $\acp >0.5$.

\begin{table}
\begin{ruledtabular}
\begin{tabular}{cccccc}
$f_K$&$B_K$&$B_K^{LR(1)}$&$B_K^{LR(2)}$&$\tilde{B}_K^{RR(1)}$
&$\tilde{B}_K^{RR(2)}$\\ \hline
159.8 MeV&0.69&1.03&0.73&0.65&1.05\\
\end{tabular}
\begin{tabular}{cccccccccc}
$f_{B_d}$&$f_{B_s}/f_{B_d}$&$B_B$&$B_{B_d,B_s}^{LR(1)}$&
$B_{B_d,B_s}^{LR(2)}$&$\tilde{B}_{B_d}^{RR(1)}$
&$\tilde{B}_{B_s}^{RR(1)}$\\ \hline
%&$\tilde{B}_{B_d}^{RR(2)}$&
%$\tilde{B}_{B_d}^{RR(2)}$\\ \hline
210 MeV&1.17&0.8&0.8&0.8&0.8&1.19\\
\end{tabular}
\end{ruledtabular}
\caption{Decay constants and bag parameters for the $B^0-\bar{B}^0$ and
the $K^0-\bar{K}^0$ mixing matrix elements used in the numerical 
calculation\cite{Yamada:2000kt}. 
$\tilde{B}_{B_d,B_s}^{RR(2)}$ are given by 
$\tilde{B}_{B_q}^{RR(2)} = 5 \tilde{B}_{B_q}^{RR(1)}
- 4 (\frac{ m_{B_q} }{ m_b + m_q })^{-2} B_{B_q}$
with $q=d,s$, which is valid in the static limit.}
\label{tab f and bag}
\end{table}

\subsection{Numerical results}

At first we discuss qualitative features of the SUSY contributions to
the \bdbd\ mixing, the \bsbs\ mixing and \ek\ for each model.

In the mSUGRA, it is well-known that the main SUSY
contributions to $M_{12}(B_d)$, $M_{12}(B_s)$, 
and $M_{12}(K)$\ come from the box diagrams with
the charginos and the up-type squarks and that the flavor mixing in 
the chargino
vertex is determined by the CKM matrix.
Consequently, the SUSY contributions are approximately proportional 
to the CKM matrix
elements $(V_{td}^*V_{tb})^2$, $(V_{ts}^*V_{tb})^2$,
and $(V_{td}^*V_{ts})^2$ for $M_{12}(B_d)$, $M_{12}(B_s)$, and
$M_{12}(K)$, respectively, and the ratios to the corresponding 
SM contributions are common:
\begin{equation}
\frac{(\dmbd)_{\rm SUSY}}{(\dmbd)_{\rm SM}}
=
\frac{(\dmbs)_{\rm SUSY}}{(\dmbs)_{\rm SM}}
\approx
  \frac{(\ek)_{\rm SUSY}}{(\ek)_{\rm SM}}.
\end{equation}
For the other two models, this proportionality is violated due to the
squark flavor mixing induced by sources other than the CKM matrix.

In the SU(5) SUSY GUT with right-handed neutrinos, the flavor mixing in
the right-handed down-type squark sector is related to the flavor mixing
in the left-handed slepton sector, and hence the constraints from the
lepton flavor violating processes such as \meg\ is important.
As shown in Ref.~\cite{Baek:2001kh}, $\br(\meg)$ exceeds the present 
experimental
upper limit $1.2\times10^{-11}$ in the parameter region where the SUSY
contributions to the \bb\ mixings become larger 
than $\simeq 10$\%\footnote{The constraint from $\br(\meg)$ is somewhat
model dependent. The above results depend on our
choice of the structure of $M_N$ and/or $V_{\text{MNS}}$.
If we change these assumptions and suppress
$\br(\meg)$, the SUSY contribution to $\dmbs$ can be more significant.
For example, if we take the small mixing MSW solution for the solar
neutrino anomaly, a 50\% enhancement of $\dmbs$ is 
possible\cite{Baek:2001kh}.
}.
Therefore \dmbd, \dmbs\ and \acp\ are almost the same as the SM values
for a given CKM matrix.
On the other hand, \ek\ can be quite different from the SM value even
under the $\mu\rightarrow e\gamma$ constraint because of
large enhancements of the $K^0-\bar{K}^0$
mixing matrix elements for the scalar operators 
in Eq.~(\ref{eq kk matrix element}).
This correction of \ek\ leads to a change of the allowed region of the
parameter $\phi_{3}$ and eventually
affects the possible region of \dmbsd\ and \acp.

In the U(2) model, 
the SUSY contribution to \ek\ can be large in a similar manner.
In addition, there are $O(1)$ corrections to $M_{12}(B_d)$ and
$M_{12}(B_s)$ so that the \dmbsd\ and \acp\ can be
different from the SM values with the same CKM matrix.

\begin{figure}
\includegraphics{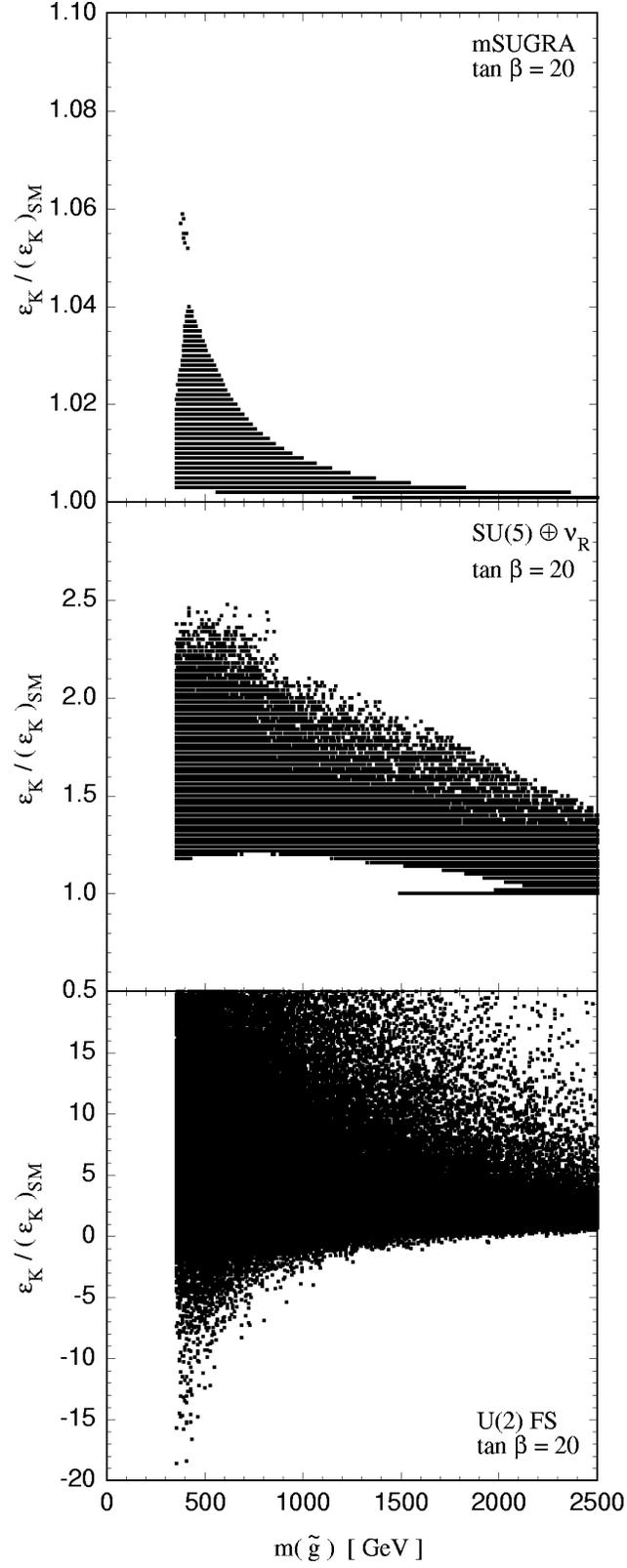}
\caption{
Ratio of \ek\ to the SM value as a function of the gluino mass for a
fixed set of the parameters in the CKM matrix (see text).
}
\label{fig:rek-gno}
\end{figure}

\begin{figure}
\includegraphics{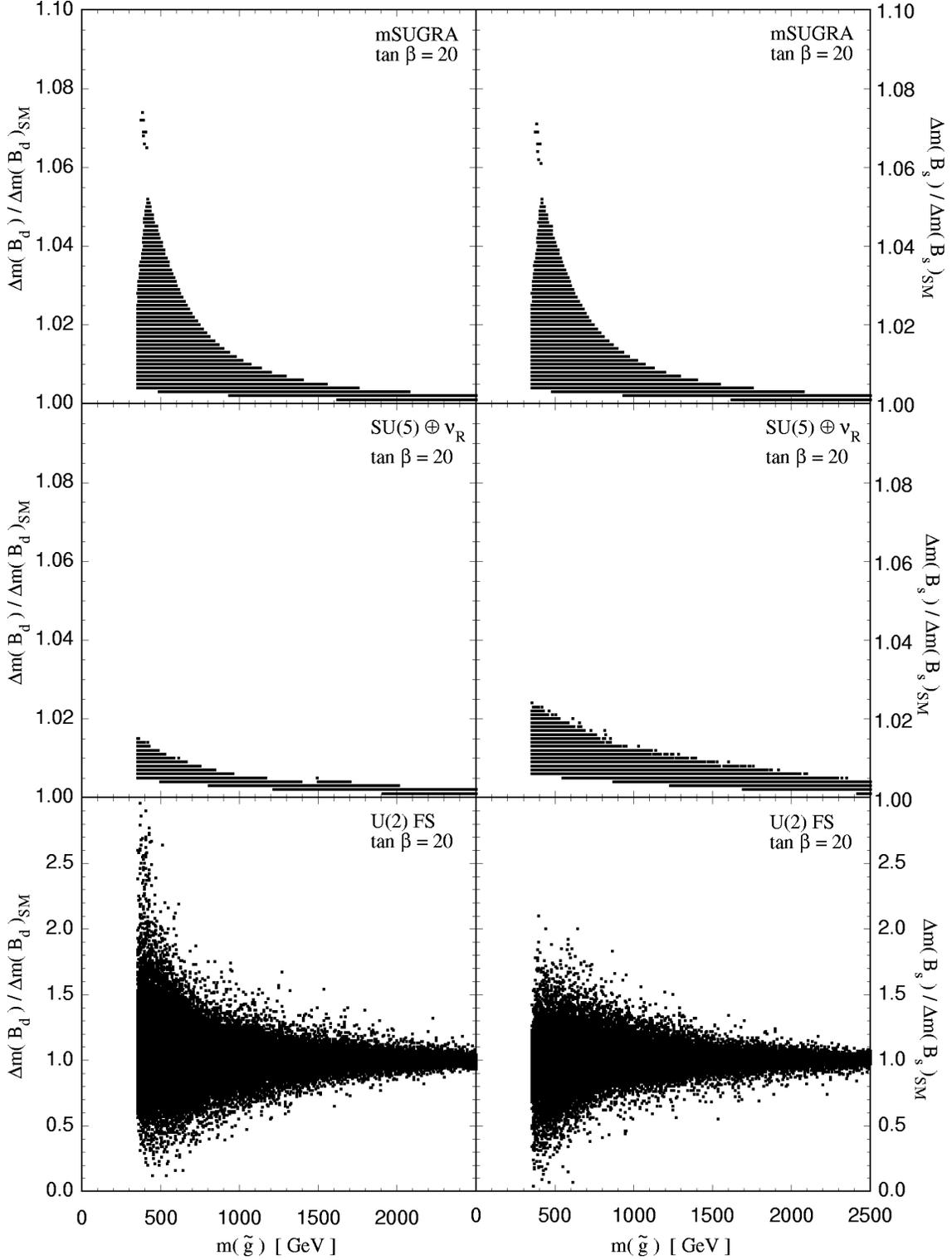}
\caption{
Deviations of \dmbd\ and \dmbs\ from the SM values as functions of the
gluino mass with the same parameter set as
Fig.~\protect\ref{fig:rek-gno}.
}
\label{fig:rdmb-gno}
\end{figure}

\begin{figure}
\includegraphics{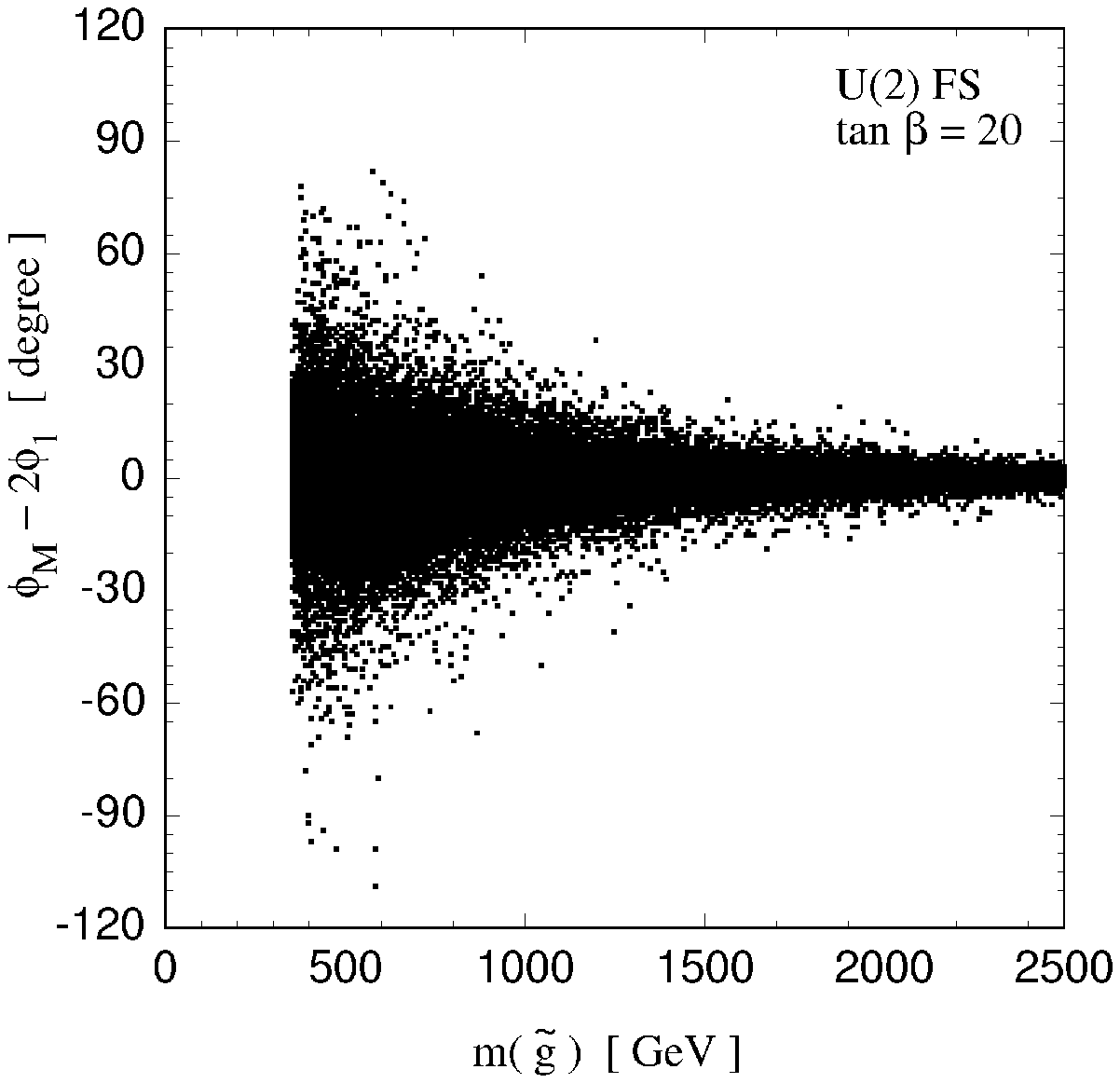}
\caption{
Deviation of $\phi_M$ from the SM value ($=2\phi_1$) as a function of
the gluino mass for the U(2) model with the same
parameter set as Fig.~\protect\ref{fig:rek-gno}.
}
\label{fig:dphiM-gno}
\end{figure}

Deviations of $\varepsilon_K$, $\Delta m_{B_d}$,
$\Delta m_{B_s}$, and $\phi_M$ from the SM
values are plotted as functions of the gluino mass
for $\tan \beta=20$ in
FIG.~\ref{fig:rek-gno}--\ref{fig:dphiM-gno}.
In these figures, we fix the parameters in the CKM matrix as
$|V_{ub}/V_{cb}|=0.09$ and $\phi_{3}=65^\circ$ and do not impose the
experimental constraints from \ek, \dmbd, \dmbs,\ and \acp.
The above features can be seen quantitatively in these figures.
We see that \ek\ can be different from the SM prediction by a factor
$\simeq 2.5$ in the SU(5) SUSY GUT with right-handed neutrinos, and the
deviation is even larger in the U(2) model.
In the mSUGRA, the deviation is smaller than $10$\%.
As for \dmbd\ and \dmbs, $O(1)$ deviations are possible in the U(2)
model, while the deviations are small for the other
cases.
The value of $\phi_M$ can also
be different from the SM value ($=2\phi_1$)
significantly in the case of the U(2) model.
For the mSUGRA and the SU(5) SUSY GUT with
right-handed neutrinos,  $\phi_M=2\phi_1$ is a good approximation.

\begin{figure}
\includegraphics{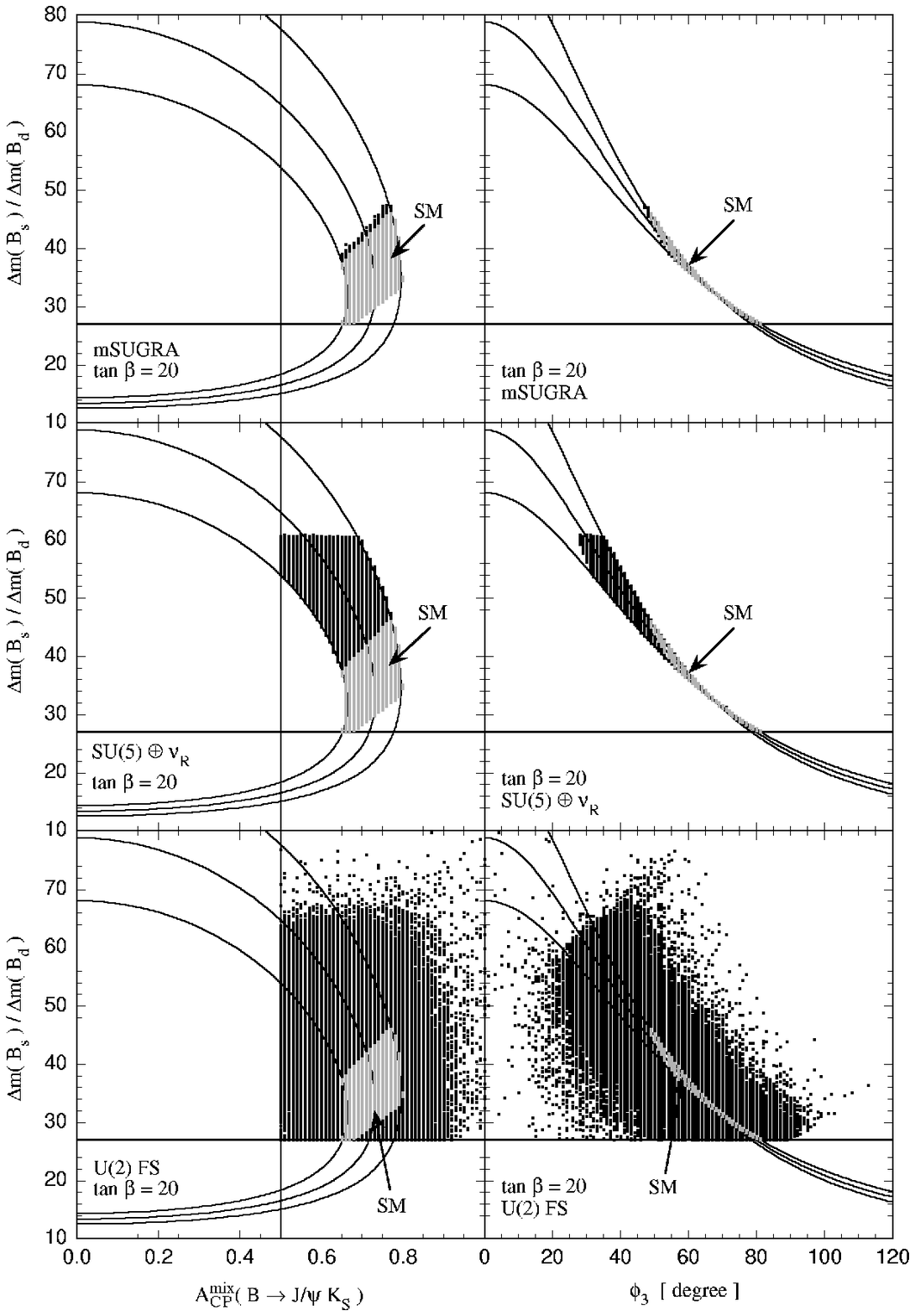}
\caption{
Scatter plots in the planes $(\acp,\,\dmbsd)$ and $(\phi_3,\,\dmbsd)$
for three SUSY models.
Solid curves show the SM values with fixed $|V_{ub}/V_{cb}|=0.08$, $0.09$
and $0.10$. 
}    
\label{fig:dmbsd-acp-phi3}
\end{figure}

Next, let us vary $|V_{ub}/V_{cb}|$ and $\phi_{3}$ and impose all the
experimental constraints explained above.
In FIG.~\ref{fig:dmbsd-acp-phi3}, we show possible values of \acp,
\dmbsd, and $\phi_3$ for the parameter sets satisfying the constraints.
The corresponding allowed region for the SM is also given in each plot.
Solid curves show the correlations among three quantities  in the SM 
for $|V_{ub}/V_{cb}|=0.08$, $0.09$,  and $0.10$. 
We see that the allowed region in the SM is mainly determined by
$|V_{ub}/V_{cb}|$ and \ek.

In the mSUGRA, the deviation from the SM is not so
significant since the SUSY contributions to all the $M_{12}$'s are
small.
 
In the SU(5) SUSY GUT with right-handed neutrinos, we see that all the
allowed points lie between the lines corresponding to the SM values
with $|V_{ub}/V_{cb}|=0.08$ and $0.10$.
This pattern arises because only the \kk\ mixing receives SUSY corrections
of $O(1)$ corrections from the SUSY loops,
whereas the SUSY contributions to  $M_{12}(B_d)$ and $M_{12}(B_s)$ are small.
As a result, the allowed region of $\phi_3$ can be shifted, and 
\dmbsd\ and \acp\ can be different from the SM region.
%This is because that the SUSY contributions to
%$M_{12}(B_d,\,B_s)$ are small.
%Although $M_{12}(B_d,\,B_s)$ is dominated by the SM contribution,
%allowed region can be different from that in the SM.
%%The difference of the allowed region from the SM comes from the change
%%of \ek.
%In a parameter region where \ek\ receives an $O(1)$ SUSY correction, the
%allowed region of $\phi_3$ shifts so that \dmbsd\ and \acp\ can be
%different from the SM region.
In other words, observables from $B$ physics namely $|V_{ub}/V_{cb}|$, 
\dmbsd, \acp, and $\phi_3$
consistently determine a set of parameters in the CKM matrix in the SM
analysis, though the experimental value \ek\ may not be consistent
with the SM value calculated by the CKM parameters from $B$ physics.
The upper limit of \dmbsd\ in the plot is determined by the
lower bound of \dmbd\ given in Eq.~(\ref{eq:dmbdrange}).

In the U(2) model, we see that the allowed
points exist outside of the region between 
$|V_{ub}/V_{cb}|=0.09\pm 0.01$ lines.
SUSY corrections to $M_{12}(B_{d})$ and $M_{12}(B_{s})$ in this model,
unlike those in the SUSY GUT, are considerably large and not 
proportional to the corresponding combinations of CKM elements 
in the SM.
Since all of \ek, \dmbsd\, and \acp\ can be corrected, there might be a
mismatch in the determination of the unitarity triangle by the SM
analysis with these quantities and $\phi_3$.
In particular, we may be able to extract new physics contributions
from observables in $B$ physics.

\begin{figure}
\includegraphics[scale=0.7]{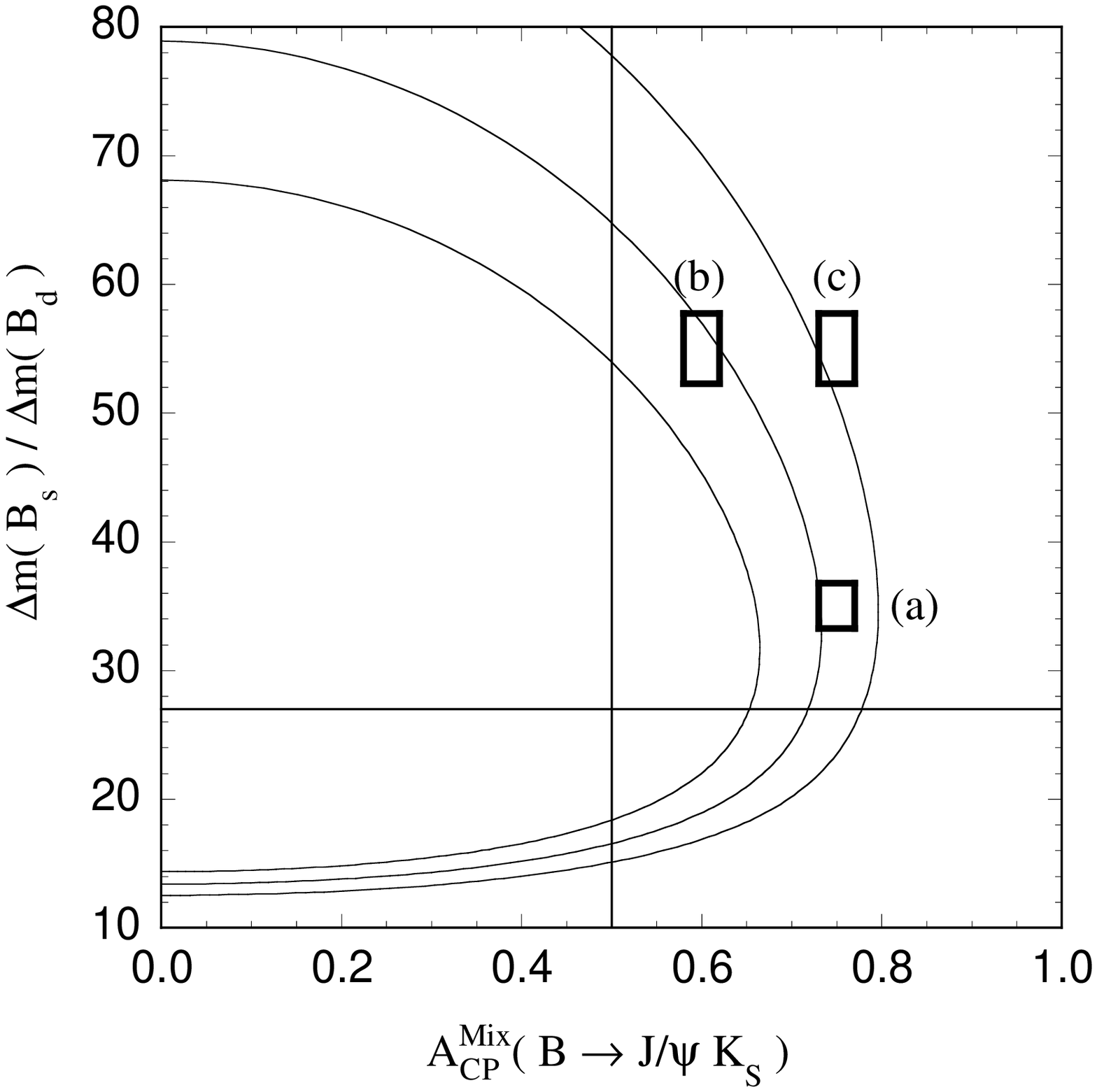}
\caption{Typical example regions. (a)$\dmbsd = 35\times (1\pm 0.05)$ and
$\acp = 0.75\pm 0.02$. (b)$\dmbsd = 55\times (1\pm 0.05)$ and
$\acp = 0.60\pm 0.02$. (c)$\dmbsd = 55\times (1\pm 0.05)$ and
$\acp = 0.75\pm 0.02$.}
\label{fig:exampleregion}
\end{figure}

Finally, we discuss future prospects of new physics search in $B$
decays.
We expect that \acp\ and \dmbsd\ will be precisely measured 
in  a few years at the $B$ factories and Tevatron experiments. 
If we assume the SM, CKM parameters, especially $\phi_3$ can be determined
from these observables with small hadronic uncertainties. By comparing this 
$\phi_3$ value with that derived from CP asymmetries in various $B$ decays,
we can carry out a consistency check of the SM 
and examine existence of SUSY effects.
As an illustration, we pick up
the calculated data points which satisfy the following values of \acp\
and \dmbsd:
\begin{displaymath}
\begin{array}{cll}
 \mbox{(a)} & \dmbsd = 35\times(1\pm0.05), & \acp = 0.75\pm0.02,
\\
 \mbox{(b)} & \dmbsd = 55\times(1\pm0.05), & \acp = 0.60\pm0.02,
\\
 \mbox{(c)} & \dmbsd = 55\times(1\pm0.05), & \acp = 0.75\pm0.02.
\end{array}
\end{displaymath}
(a) corresponds to the case in which \ek, \acp, and \dmbsd\ are consistent
with the SM.  (b) and (c) are cases in which there are some inconsistencies
among the three observables within the SM.
The three regions are shown in FIG.~\ref{fig:exampleregion}.

\begin{figure}
\includegraphics{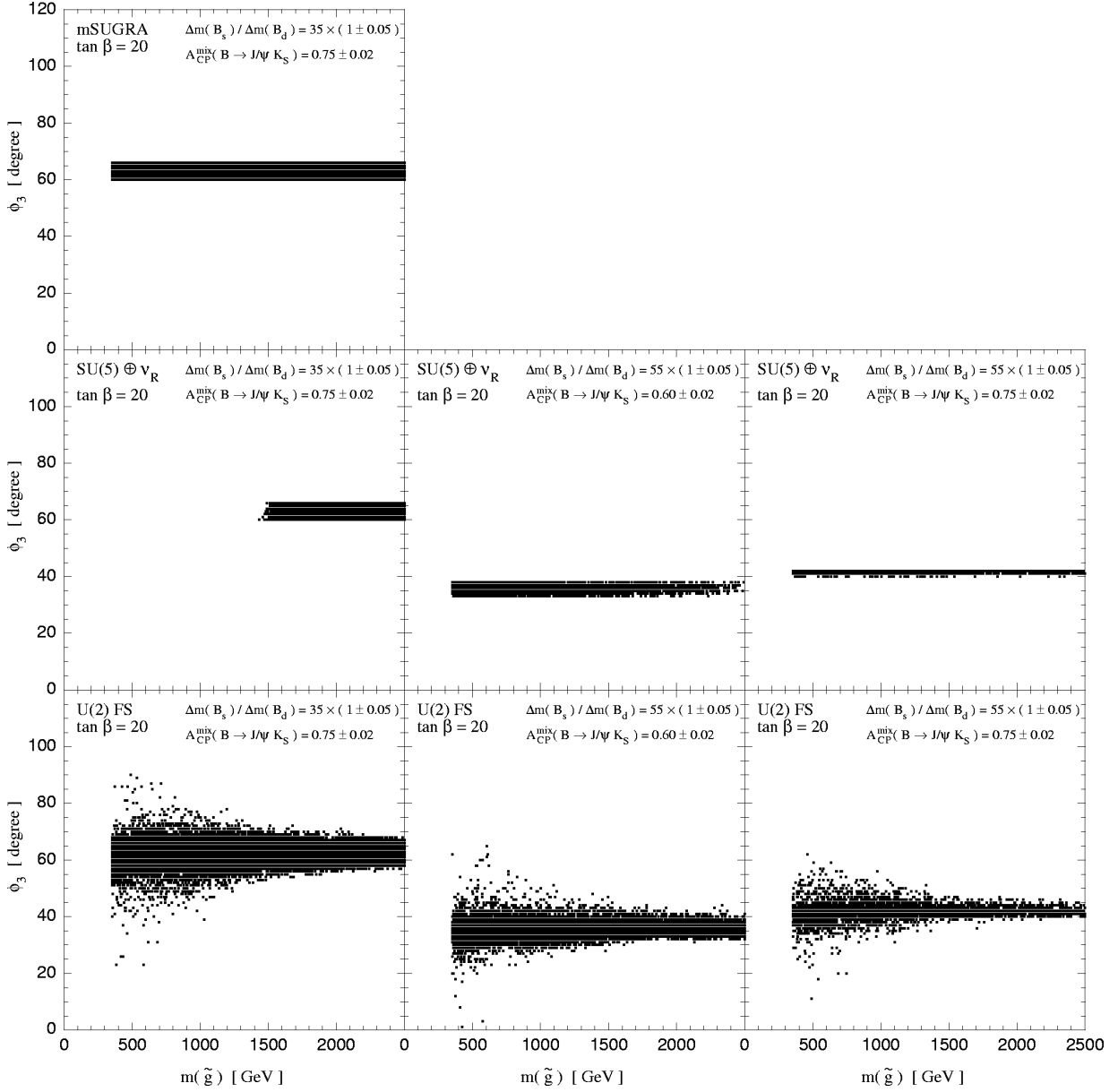}
\caption{
Possible region of $\phi_3$ as a function of the gluino mass.
}    
\label{fig:phi3-gno}
\end{figure}

We present the possible region of $\phi_3$ in each case for the three
models in FIG.~\ref{fig:phi3-gno}. 
For the case (a), $\phi_3$ is 60$^\circ$--65$^\circ$ if we assume the 
SM. The possible value of $\phi_3$ is the same as the SM 
in the mSUGRA and the SU(5) SUSY GUT with right-handed neutrinos.
The parameter region with $m_{\tilde{g}}\alt1.5$ TeV in the SU(5)
SUSY GUT is excluded due to the \meg\ constraint. 
On the other hand, in the U(2) model, $\phi_3$ can be
different from the SM value by $\sim30^\circ$ for
$m_{\tilde{g}}\alt1$ TeV region.

For the cases (b) and (c), the mSUGRA (as well as the SM) is
excluded because of the mismatch among \ek, \acp, and \dmbsd.
In the other two models, the experimental value of \ek\ can be reproduced
with SUSY contributions.
In the SU(5) SUSY GUT with right-handed neutrinos, 
$\phi_3$ is the same as that derived from  \dmbsd and \acp\
by the SM analysis.
In the U(2) model, $\phi_3$ can be different from the value of the SM.

We have also studied the case of $\tan\beta=5$ 
and drawn the figures corresponding to
FIG.~\ref{fig:dmbsd-acp-phi3} and \ref{fig:phi3-gno}.
We have found that the allowed regions in these figures are almost 
same as those for $\tan\beta=20$.

%%%%%%%%%%%%%%%%%%%%%%%%%%%%%%%%%%%%%%%%%%%%%%
% Section 5 of the paper by T.Goto, Y.Okada, %
% Y.Shimizu, T.Shindou and M.Tanaka.         %
% Written by T.S.			     %
%%%%%%%%%%%%%%%%%%%%%%%%%%%%%%%%%%%%%%%%%%%%%%
\section{Conclusions\label{CONCLUSIONS}}
In order to distinguish SUSY models by measurements at
$B$ factories,
we have studied SUSY contributions to the \kk, \bdbd, and \bsbs\ mixings
in three SUSY models, namely the mSUGRA, the SU(5) SUSY GUT 
with right-handed neutrinos, and the U(2) model.

First, we have considered the observables $\dmbd$, $\dmbs$, $\acp$, 
and $\ek$.
In the mSUGRA, the deviations from the SM values are at most 10 percents 
for these observables.
In the SU(5) SUSY GUT with right-handed neutrinos,
the SUSY contributions to $\ek$ can be large whereas
those to $M_{12}(B_d)$ and $M_{12}(B_s)$ are less than 10\%.
In the U(2) model, the deviations from the SM values for all the above
observables can be very large.
In the latter two models, we may be able to see SUSY effects from
the consistency check of the unitarity triangle among \ek,
$\dmbs/\dmbd$, and $\acp$.

Second, we have considered cases in which the two observables $\dmbsd$ 
and $\acp$ are precisely determined at the $B$ factories and Tevatron
experiments. We have studied how we can distinguish these different models 
if we determine $\phi_3$ in addition to the above two observables.
We can carry out the consistency check of the unitarity triangle
among the observables in $B$ physics, namely $\dmbs/\dmbd$, and $\acp$,
 and $\phi_3$. For the U(2) model, in particular, a large deviation
from the SM value is possible. It is therefore very important
to determine $\phi_3$ precisely in theoretically clean ways from 
the decay modes, such as 
$B\rightarrow \pi\pi,\rho\pi,D^{(*)}K^{(*)},D^{(*)}\pi,D^{*}\rho$.
These measurements are possible in future $e^+e^-$ super $B$ factories 
and hadron machines such as LHC-B and B-TeV.

In this paper, we have mainly considered the consistency test of the 
unitarity triangle through $B_d$ decays,
but there are other possibilities of finding SUSY effects in $B$ physics.
A new phase in the \bsbs\ mixing amplitude may affect CP asymmetries 
in $B_s$ decays such as the $B_s\to J/\psi\,\phi$ decay. 
These asymmetries can be measured in $B$ experiments at hadron
machines. For the U(2) model, these CP asymmetries could be
different from the SM prediction\cite{Masiero:2001cc}.
Another possibility is to measure branching ratios
and CP asymmetries in rare decays such as $b\to s l^+ l^-$
and $b\to s\nu\overline{\nu}$.

In conclusion we have shown that SUSY models with different flavor
structures can be distinguished in $B$ physics.  
As we have illustrated with three specific models,
the patterns of the deviations from the SM in the $B$ physics depend on
the SUSY breaking mechanism and interactions at a high energy scale.
Present and future experiments in $B$ physics at $e^+e^-$ $B$ factories 
and hadron machines are therefore very important to explore
flavor structure of SUSY breakings.

%%%%%%%%%%%%%%%%%%%%%%%%%%%%%%%%%%%%%%%%%%%%%%%
% Acknowledgements
%%%%%%%%%%%%%%%%%%%%%%%%%%%%%%%%%%%%%%%%%%%%%%%
\acknowledgments{
This work was supported in part by a Grant-in-Aid of 
the Ministry of Education, Culture, Sports, Science and
Technology, Government of Japan(No.~12440073).
The work of Y.O. was supported
in part by a Grant-in-Aid of the Ministry of Education, Culture, Sports,
Science, and Technology, Government of Japan(No.~13640309),
priority area ``Supersymmetry and 
Unified Theory of Elementary Particles''(No.~707).
The work of Y.S. was supported in part by a Grant-in-Aid of 
the Ministry of Education, Culture, Sports, Science and
Technology, Government of Japan(No.~13001292).
}
\section*{Note added}
After submission of this paper, we received the paper by D. Chang, 
A. Masiero, and H. Murayama\cite{Chang:2002mq}
in which possibility of the large $b$--$s$ transition
is pointed out in the context of the SO(10) SUSY GUT.

%%%%%%%%%%%%%%%%%%%%%%%%%%%%%%%%%%%%%%%%%%%%%%
% Appendix of the paper by T.Goto, Y.Okada,  %
% Y.Shimizu, T.Shindou and M.Tanaka.         %
% Written by T.G.			     %
%%%%%%%%%%%%%%%%%%%%%%%%%%%%%%%%%%%%%%%%%%%%%%
\appendix*

\section{The CKM matrix and the unitarity triangle}
\label{app:CKM}
In this paper, we use the ``standard'' parameterization\cite{Chau:1984fp}
for the CKM
matrix with three mixing angles $\theta_{12}$, $\theta_{23}$,
$\theta_{13}$ and a complex phase $\delta_{13}$:
\begin{equation}
V_{\rm CKM} =
  \left(
    \begin{array}{ccc}
       c_{12}c_{13} & s_{12}c_{13} & s_{13}\e^{-i\delta_{13}} \\
      -s_{12}c_{23}-c_{12}s_{23}s_{13}\e^{i\delta_{13}} &
       c_{12}c_{23}-s_{12}s_{23}s_{13}\e^{i\delta_{13}} & s_{23}c_{13} \\
       s_{12}s_{23}-c_{12}c_{23}s_{13}\e^{i\delta_{13}} &
      -c_{12}s_{23}-s_{12}c_{23}s_{13}\e^{i\delta_{13}} & c_{23}c_{13}
    \end{array}
  \right),
\label{eq:ckm-pdg}
\end{equation}
where $c_{ij} = \cos\theta_{ij}$ and $s_{ij} = \sin\theta_{ij}$.
The angles $\phi_1$, $\phi_2$ and $\phi_3$ in the unitarity triangle are
defined as
\begin{subequations}
\begin{eqnarray}
  \phi_1 &=& \arg\left(-\frac{V_{cb}^*V_{cd}}{V_{tb}^*V_{td}}\right),
\\
  \phi_2 &=& \arg\left(-\frac{V_{tb}^*V_{td}}{V_{ub}^*V_{ud}}\right),
\\
  \phi_3 &=& \arg\left(-\frac{V_{ub}^*V_{ud}}{V_{cb}^*V_{cd}}\right).
\end{eqnarray}
\end{subequations}
In the convention (\ref{eq:ckm-pdg}), these angles are written 
in a good approximation as
\begin{eqnarray}
  \phi_1 &=& \frac{1}{2}\arg M_{12}^{\rm SM}(B_d),
\\
  \phi_3 &=& \delta_{13}.
\end{eqnarray}

%%%%%%%%%%%%%%%%%%%%%%%%%%%%%%%%%%%%%%%%%%%%%%
% References of the paper by T.Goto, Y.Okada,%
% Y.Shimizu, T.Shindou and M.Tanaka.         %
% Written by T.S.			     %
%%%%%%%%%%%%%%%%%%%%%%%%%%%%%%%%%%%%%%%%%%%%%%
 
\end{document}